\documentclass{aa}
\include{epsf}

\begin{document}

\title{Numerical simulations of kinematic dynamo action }

\author{V. Archontis \inst{1}
\and S.B.F. Dorch\inst{1,2}
\and {\AA}. Nordlund\inst{1}
}

\offprints{V.D. Archontis --- bill@astro.ku.dk}

\institute{
     The Niels Bohr Institute for Astronomy, Physics and Geophysics,
     Juliane Maries Vej 30, DK-2100 Copenhagen {\O}, Denmark
\and
     The Institute for Solar Physics of the
     Royal Swedish Academy of Sciences,
     SCFAB, SE-10691 Stockholm, Sweden
}

\date{Received date, accepted date}

\authorrunning{Archontis et al.}

\def\Rm{{\rm Re}_{\rm m}}
\def\Ekin{{\rm E}_{\rm kin}}
\def\Emag{{\rm E}_{\rm mag}}
\def\Etot{{\rm E}_{\rm tot}}
\def\Qvisc{{\rm Q}_{\rm v}}
\def\Qjoule{{\rm Q}_{\rm J}}
\def\Qcool{{\rm Q}_{\rm cool}}
\def\Wlor{{\rm W}_{\rm L}}
\def\uu{{\bf u}}
\def\bb{{\bf B}}
\def\jj{{\bf j}}
\def\ee{{\bf e}}
\def\BB{{\bf B}}
\def\BE{\begin{equation}}
\def\EE{\end{equation}}
\def\div{\nabla \cdot}

\abstract{Numerical simulations of kinematic dynamo action in
steady and three-dimensional ABC flows are presented with special
focus on the difference in growth rates between cases with single
and multiple periods of the prescribed velocity field. It is found
that the difference in growth rate (apart from a trivial factor
stemming from a scaling of the rate of strain with the wavenumber
of the velocity field) is due to differences in the recycling of
the weakest part of the magnetic field. The single wavelength
classical ABC-flow experiments impose stronger symmetry
requirements, which results in a suppression of the growth rate.
The experiments with larger wave number achieve growth rates that
are more compatible with the turn-over time scale, by breaking the
symmetry of the resulting dynamo generated magnetic field.
Differences in the topology in cases with and without stagnation
points in the imposed velocity field are also investigated, and it
is found that the cigar-like structures that develop in the
classical A=B=C dynamos, are replaced by ribbon structures in
cases where the flow is without stagnation points.
}

\maketitle

\section{Introduction}
Dynamo theory aims at explaining how magnetic fields can be
generated and sustained by fluid motions, in the presence of
dissipation. Studies of dynamo action in astrophysical objects are
usually connected with the question of the existence of fast
dynamos (Vainshtein \& Zeldovich \cite{Vainshtein+Zeldovich1972}).
The fast dynamo problem refers to the generation of magnetic fields
in the limit of infinite magnetic
Reynolds number ${\rm Re}_{\rm m}$, and has been extensively
reviewed by e.g.\ Childress \& Gilbert
(\cite{Childress+Gilbert1995}).

In the simplest formulation of the
kinematic dynamo problem, an initially weak magnetic field is
amplified due to the motions of a prescribed, stationary velocity
field. Flows that have chaotic transport properties are prime
candidates for fast dynamos. They have the ability to amplify the magnetic field
exponentially by stretching the magnetic field lines. One such
class of flows that is often invoked in fast dynamos studies is
the ABC flows (named after Arnold, Beltrami \& Childress). A first
detailed report of the kinematic dynamo action of ABC flows may be
found in Dombre et al.\ (\cite{Dombre+ea86}), who examined
especially the internal symmetries of the velocity field. There
is, however, no guarantee that such symmetries are preserved in
the eigenmodes of the magnetic field. Some discussion of possible
symmetry breaking in the eigenmodes may be found in Galloway et
al.\ (\cite{Galloway+ea86}) and in Galanti et al.\ (\cite{Galanti+ea93}).
They found no evidence of symmetry breaking when the magnetic
Reynolds number is limited to very low values, although
symmetries are broken for higher values of the magnetic Reynolds
number. The presence of symmetry breaking was also shown by
Galloway et al.\ (\cite{Galloway+ea86}) when the wavenumber of the
periodic ABC flow is equal to 3 and 4. The conclusion was that the
physical reason for the symmetry breaking is unclear and
requires further investigation.

Dynamo action in ABC-flows has also been studied to determine the growth rate
of the magnetic field for different values of the magnetic
Reynolds number (Arnold et al.\ \cite{Arnold+ea83}; Galloway et
al.\ \cite{Galloway+ea86}; Lau \& Finn \cite{Lau+Finn93}). Results
for the case with unit wavenumber, for magnetic Reynolds number
up to ${\rm Re}_{\rm m}=1600$ (Dorch \cite{Dorch2000}) showed that
the growth rate saturates at a value close to 0.07.
Galanti et al.\ (\cite{Galanti+ea93}) found
that doubling the wavenumber of the flow
leads to a sharp increase of the growth rate, while
for larger wavenumbers the growth rate scales with the strain rate
of the flow (i.e., in proportion to the wave number).

In the current paper we address some of the issues mentioned above by
means of numerical kinematic MHD simulations. The dynamics of the
magnetic structures in physical space is examined in connection
with the amplification mechanism which drives the dynamo action in
ABC flows. We also investigate the physical reason that
causes the increase of the growth rate of an ABC flow for
wavenumbers larger than unity. To achieve this we follow the
motion of magnetic field lines in time and demonstrate that the
transportation of the weak part of the field is a key factor of
the dynamo process.

The numerical aspects of the simulations are
described in the following section. The growth rates produced by
fast kinematic dynamo action of ABC flows are presented in Section
\ref{rates.sec}. Visualizations of the magnetic structures and
field lines in physical space are shown in Section
\ref{structures.sec}. Dynamo action produced by flows without
stagnation points is studied in Section \ref{points.sec}. Section
\ref{conclusions.sec} contains the overall conclusions of the
numerical simulations presented here.

\section{The simulation}
\label{simulation.sec}

Our aim is to study kinematic dynamo action in ABC flows. The time
evolution of {\bf B} is governed by the (dimensionless) induction
equation:
\begin{equation}
\frac{\partial {\bf B}}{\partial t} = \nabla \times ( {\bf u}
\times {\bf B})
    + \frac{1}{{\rm Re}_{\rm m}} \nabla^2 {\bf B}, \label{induction.eq}
\end{equation}
\begin{equation}
\nabla\cdot{\bf B}=\nabla\cdot{\bf u}=0, \label{solenoidal.eq}
\end{equation}
The velocity field is chosen to have the form of an ABC flow:
\begin{eqnarray}
{\bf u}_{\rm ABC} & = & A(0,\sin kx, \cos kx) + B(\cos ky,0,\sin
ky)\\ \label{abcflow.eq}
    & +  & C(\sin kz,\cos kz,0) ~, \nonumber
\end{eqnarray}
where $k$ is the wavenumber of the flow.
The special case with parameters A=B=C=$k$=1 is referred to as the
``normal" ABC flow dynamo. In the normal case there are two
windows of dynamo action (where the growth rates are positive): A
closed one in the interval Re$_{\rm m}$=8.9--17.5, and an open one
for magnetic Reynolds numbers above 27 (Galloway et al.\
\cite{Galloway+ea84}).

Eq.\ (\ref{induction.eq}) is solved numerically on a staggered
mesh using array valued functions for evaluating the space and
time derivatives. A third-order predictor-corrector method (Hyman
\cite{Hyman1979}) is used for the time-stepping. The applied
numerical scheme (by Galsgaard, Nordlund and others, see Galsgaard
\& Nordlund \cite{Galsgaard+Nordlund1997}; Nordlund et al.\
\cite{Nordlund+ea92}) ensures Eq.\ (\ref{solenoidal.eq}).
Periodicity of $2\pi$ is assumed over a three-dimensional
computational domain. The initial magnetic field is a random
perturbation with an amplitude of $10^{-5}$.

\section{Growth rates}
\label{rates.sec}

Previous numerical simulations (e.g.\ Arnold et al.\
\cite{Arnold+ea83}; Galloway et al.\ \cite{Galloway+ea84}; Galanti
et al.\ \cite{Galanti+ea93}; Dorch \cite{Dorch2000}) have shown
that the normal (A=B=C=$k$=1) ABC flow very likely acts as a
fast dynamo with a growth rate close to 0.077 (corresponding to a
time-scale of $\approx 0.25$ turn-over times). When the wavenumber
$k$ is larger than unity (where $k=1$ corresponds to the largest
scale of $2\pi$
in the computational domain) the growth rate scales approximately
with the rate of strain. These results are confirmed in our
simulations and we continue the study of the growth rate of the
magnetic field for higher values of ${\rm Re}_{\rm m}$.

\begin{figure}[htb]
\centering \makebox[8cm]{ \epsfxsize=7.0cm \epsfysize=6.0cm
\epsfbox{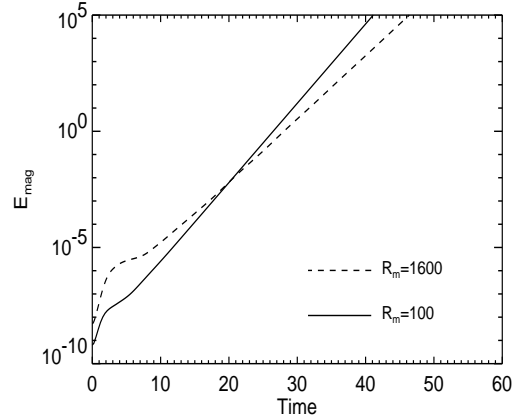}}
\caption[]{\small Temporal evolution of the magnetic energy for
the kinematic ABC dynamo with A=B=C=1 and $k=2$ for ${\rm Re}_{\rm
m}=100$ and ${\rm Re}_{\rm m}=1600$.}\label{fig1}
\end{figure}

We first discuss the case with A=B=C=1 and $k=2$ before turning
our attention to higher wavenumbers: As shown in Fig.\
(\ref{fig1}) the magnetic field is amplified exponentially
without oscillations. Initially, several modes may be present,
associated with oscillatory growth of the magnetic field (as in
the case of a normal ABC flow, Dorch \cite{Dorch2000}), but
eventually the mode with the highest growth rate is the only one
that remains in the solution (i.e.\ the dominant mode).

\begin{figure}[htb]
\centering \makebox[8cm]{ \epsfxsize=7.0cm \epsfysize=6.0cm
\epsfbox{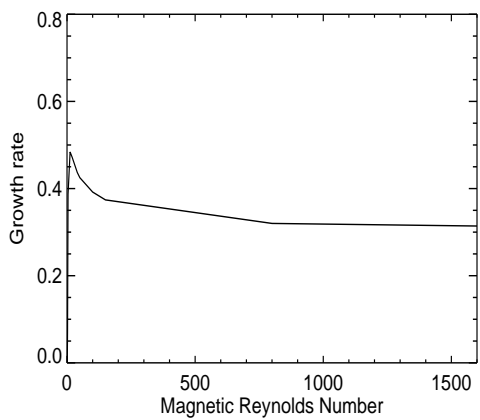}} \caption[]{\small
Growth rate of the magnetic energy (normalized by the rate of
strain) versus the magnetic Reynolds number for the kinematic ABC
dynamo with $A=B=C=1$ and $k=2$.} \label{fig2}
\end{figure}

For low ${\rm Re}_{\rm m}$ the growth rate increases with $\Rm$
and reaches its maximum value at ${\rm Re}_{\rm m}=12$. This value of
the magnetic Reynolds number belongs to the first dynamo window
for the normal ABC flow dynamo, and corresponds to a maximum
growth rate of the magnetic field in this window (Galanti et al.\
\cite{Galanti+ea93}). With $k=2$ the lowest $\Rm$ possible for
dynamo action is found to be approximately equal to 2. For large ${\rm Re}_{\rm
m}$ the growth rate decreases slightly (in contrast to the low
${\rm Re}_{\rm m}$ results of Galanti et al.\ \cite{Galanti+ea93})
but always remains positive and bounded away from zero (Fig.\
\ref{fig2}). The largest value of $\Rm$ considered here is
${\rm Re}_{\rm m}=1600$. In the high ${\rm Re}_{\rm m}$ regime the
growth rate saturates at a value close to 0.3.

The evidence indicates that the ABC flow  with $k=2$ is a fast
dynamo, since the growth rate seems to saturate at a non-zero value for high
${\rm Re}_{\rm m}$. The reason for the enhancement of the growth
rate when $k=2$ is discussed in the following section.

\begin{figure*}[htb]
\centering
\makebox[16cm]{
\epsfxsize=7.0cm
\epsfysize=6.0cm
\epsfbox{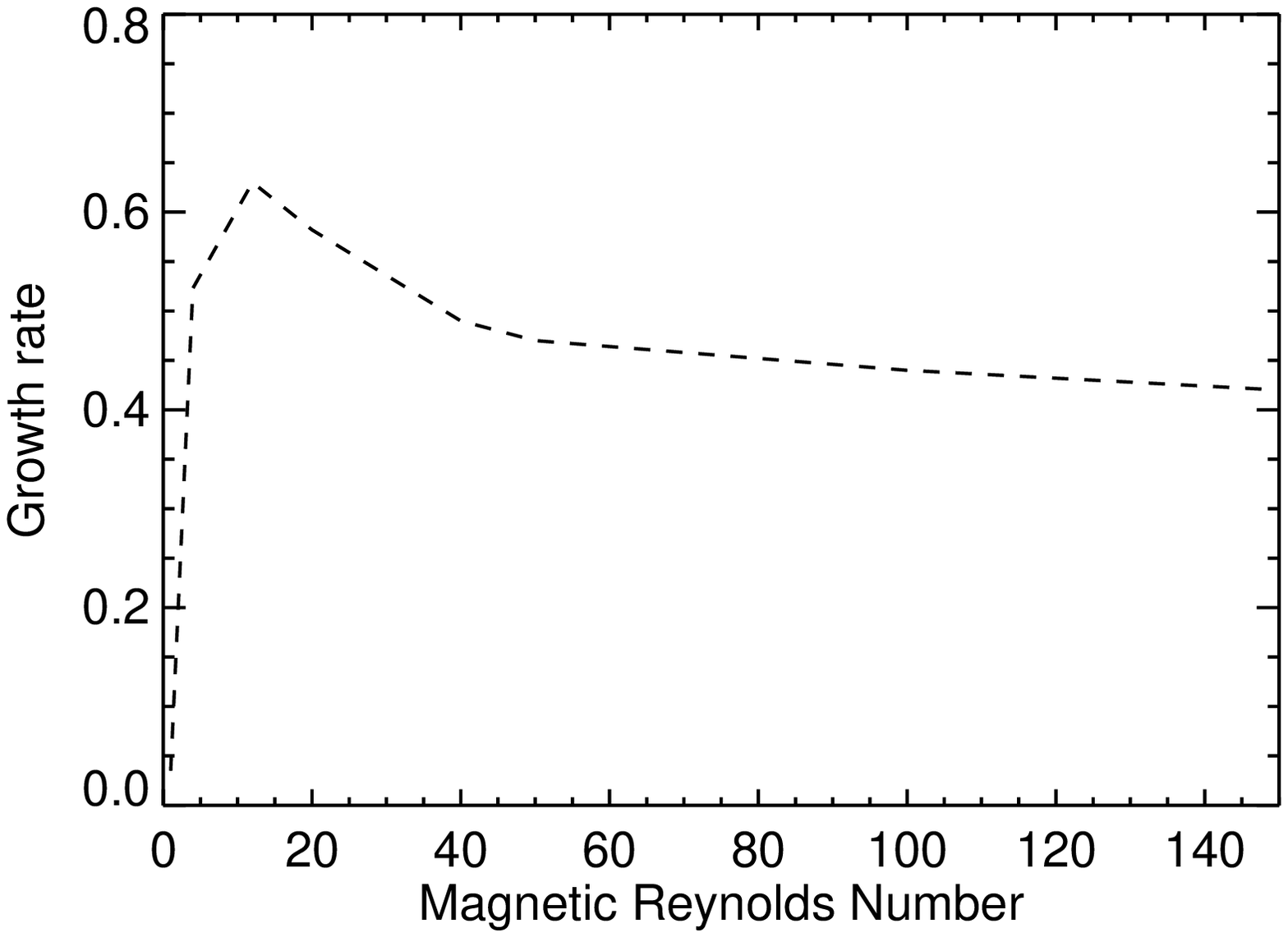}
\epsfxsize=7.0cm
\epsfysize=6.0cm
\epsfbox{fig3b.ps} }
\caption[]{\small Two panels showing the growth rate of the
magnetic energy normalized by the rate of strain for the kinematic
ABC dynamo with $k=3$ (left panel) and $k=4$ (right panel).}
\label{fig3}
\end{figure*}

The results for the growth rate of the magnetic energy at
different values of magnetic Reynolds number when $k=3$ and $k=4$
are shown in (Fig.\ \ref{fig3}).
A comparison of these results with the $k=2$ case (Fig.\
\ref{fig2}) shows that the growth rate of the magnetic
field increases with increasing wave number, in approximate
proportionality to $k$ for $k \ge 2$, scaling approximately
as $0.15\times k$, while the growth rate
is unproportionally small ($\approx 0.07$) for $k=1$.

\section{Magnetic structures}
\label{structures.sec}

\begin{figure}[!htb]
\centering \makebox[8cm]{ \epsfxsize=8.0cm \epsfysize=8.0cm
\epsfbox{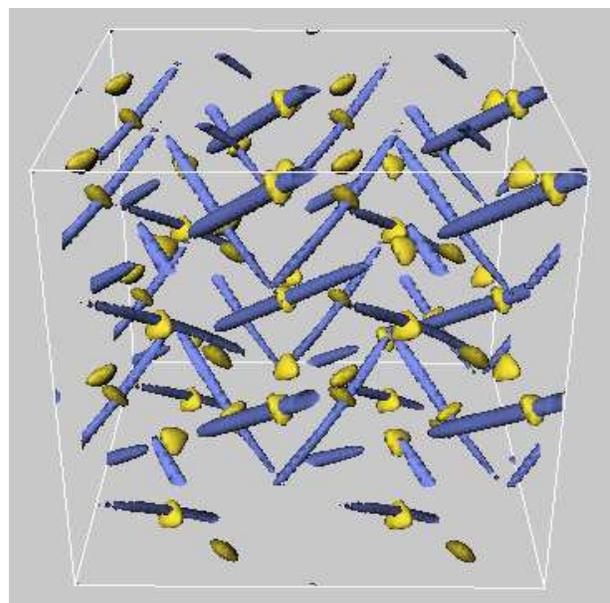} } \caption[]{\small Magnetic field
strength isosurfaces (dark) and stagnation points of the flow
(light). The magnetic field is concentrated along flux cigars
centered at the $\alpha$-type stagnation points of the flow. The
stagnation points with no flux cigars centered at them are the
$\beta$-type points. In each cell three cigars form a triangle
around a $\beta$-type stagnation point and another one is pointing
towards it. The magnetic field strength of the isosurfaces is
$70\%$ of the peak value in the snapshot.}
\label{fig4}
\end{figure}

An experiment with ${\rm Re}_{\rm m}=100$ and a numerical
resolution of $64^{3}$ was performed in order to study the
dynamics of the magnetic structures in physical space when the
wavenumber $k$ of the flow is equal to 2. The computational domain
then contains eight periodic cells, where each one (initially)
reflects the geometrical properties of a standard ABC flow. The
magnetic field evolves according to the induction equation Eq.\
(\ref{induction.eq}) and flux ``cigars" are formed at the
so-called $\alpha$-type stagnation points (Dombre et al.\
\cite{Dombre+ea86}). These are points (stable manifolds of the
flow) with converging streamlines in the plane of the stagnation
point and the field is rapidly advected to these regions (Fig.\
\ref{fig4}). The flux cigars are aligned along the axis of
divergence through the $\alpha$-type stagnation points.

At the $\beta$-type stagnation points (unstable manifolds of the
flow) the field is weaker, and takes the form of flux sheets on
both sides of the plane of divergence. The
flux sheets are formed because of the converging flow along the
axis of divergence from the $\alpha$-type stagnation points. The
flux cigars thus appear to be connected by flux sheets through
heteroclinic stream lines.

\begin{figure}[htb]
\centering
\makebox[8cm]{
\epsfxsize=8.0cm
\epsfysize=8.0cm
\epsfbox{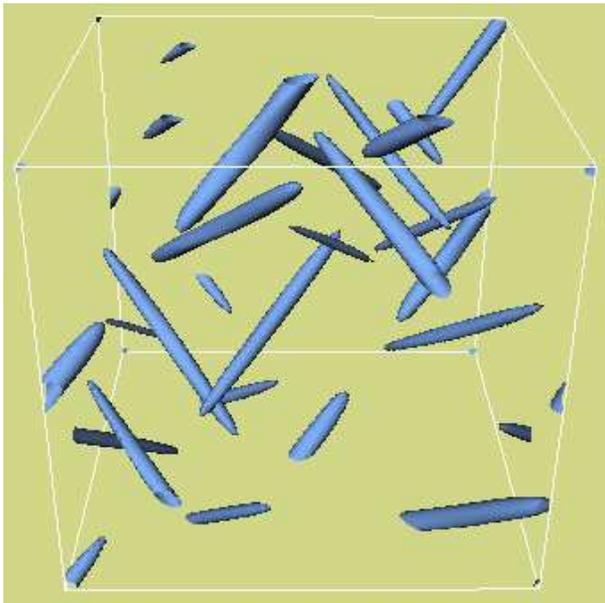} } \caption[]{\small A
snapshot showing the structure of the magnetic field when the
fastest growing mode appears and the symmetry is not conserved.
Each cell consists of flux cigars which have different size and
different field strength.} \label{fig5}
\end{figure}

When the symmetry breaking appears, and the fastest growing
eigenmode takes over, the topology of the magnetic field changes.
The fastest growing mode does not follow the symmetric
configuration of a standard ABC flow---the strong magnetic field is
still elongated into flux ``cigars" but their size and field
strength vary from cell to cell (Fig.\ \ref{fig5}).

\begin{figure}[!htb]
\centering
\makebox[8cm]{
\epsfxsize=8.0cm \epsfysize=7.0cm \epsfbox{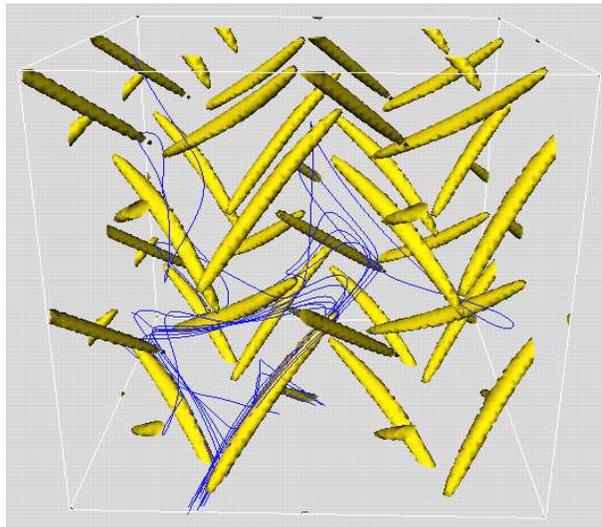}
}
\caption[]{\small
Visualization of magnetic field lines and flux cigars when the
dominant mode starts to take over. The stretching of the field
lines at the corner of a triangle at the bottom-left cell is
obvious. These field lines get pressed up against another cigar in
a neighboring cell (up-right).} \label{fig6}
\end{figure}

\begin{figure}[!htb]
\makebox[8cm]{ \epsfxsize=8.0cm \epsfysize=8.0cm
\epsfbox{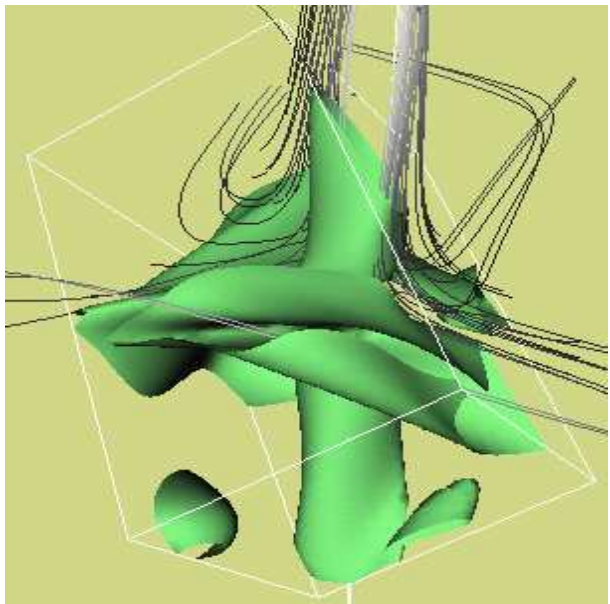} } \caption[]{\small Folding
of week flux sheets at the plane of a $\beta$-type stagnation point. The
field lines (top of the figure) are coming from a neighboring flux
cigar and change direction across the plane.} \label{fig7}
\end{figure}

The rate of growth (and decay) of individual flux cigars is
different, rather than being uniform as in the normal $k=1$ case.
The magnetic field for the latter case is recycled inside its cell
because of the exact symmetry. The time evolution of the magnetic
energy follows a periodic (oscillatory) behavior which is
associated with the conservation of symmetry in the fastest
growing eigenmode (Dorch \cite{Dorch2000}) in contrast to the
dominant mode for the $k=2$ case which is a symmetry breaking mode
(Galloway et al.\ \cite{Galloway+ea86}). Visualization of magnetic
field lines shows that they are not ``trapped" in each individual
cell but there is always a ``loose" weak field which is
transported from cell to cell and is stretched up against the
local flux cigars that are passed on the way (Fig.\
\ref{fig6}).
In general, the increase of the magnetic energy is associated with
the stretching, twisting and folding of the magnetic field lines
against the local flux cigars in the $\beta$-type planes. The
field lines coming from a flux cigar are connected through the
weak field to the cigars that form the triangles around the
$\beta$-type stagnation points. The field lines at the corners of
the triangle are stretched and added to other flux cigars of
neighboring cells through the diverging stream lines. The flux
sheets that are formed on both sides of the $\beta$-type plane
are also folded constructively. The $\beta$-type plane is actually
a plane of discontinuity where
the magnetic field lines change direction and are released to move
back into the general flow (Fig.\ \ref{fig7}).

\section{Flows without stagnation points}
\label{points.sec}

In view of the r\^{o}le played by the stagnation points in the
normal ABC flow dynamo, it is natural to ask whether they are a
necessary ingredient for fast dynamo action. When the parameters
A, B and C are varied it is possible for the stagnation points to
be created or destroyed (Dombre et at.\ \cite{Dombre+ea86}).
When A=5, B=2, C=2 or A=5, B=2, C=1 (abbreviated as the $5:2:2$
and $5:2:1$ case, respectively) the flow has no stagnation points.
Numerical simulations were performed for both cases and the
results obtained are similar. Therefore, only the former case is
discussed in detail.

\begin{figure}[!htb]
\centering
\makebox[8cm]{
\epsfxsize=8.0cm \epsfysize=8.0cm \epsfbox{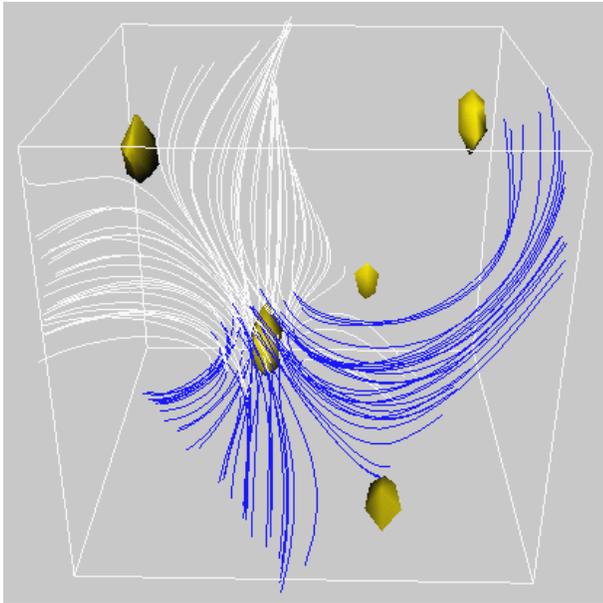}
}
\caption[]{\small The bead-shaped
isosurfaces show the position of the points with minimum velocity
of the $5:2:2$ flow. Visualization of stream lines of the flow
shows that they diverge when they reach the neighborhood of other
points with minimum velocity.} \label{fig8}
\end{figure}

The $5:2:2$ flow has four points where the velocity is close to
zero (but not zero, hence there are no stagnation points). The
disappearance of the stagnation points means that there are no
longer any heteroclinic stream lines that connect the points of minimum
velocity. However, there are stream lines that converge to
these points and stream lines that diverge from these points
(Fig.\ \ref{fig8}). The stream lines of the $5:2:2$
flow do not follow any specific symmetry as in the normal ABC flow
(which has a three-fold symmetry of stream lines around the
stagnation points) and the topology of the flow becomes more
complex.

The importance of these minimum velocity points is that they
coincide with local extrema of the ``perpendicular convergence" of
the local motion to a magnetic field line
\begin{equation}
D_{\perp} \ln {\bf B}/Dt = -\nabla\cdot{\bf u_{\perp}}
\label{convergence.eq}
\end{equation}
which, apart from diffusive effects, gives the exponential rate of
increase of the local magnetic field amplitude (Nordlund et al.\
\cite{Nordlund+ea94}). In general, it is the gradients of the flow
velocity that enable the stretching of field lines that is so essential to
the dynamo mechanism, and consequently these points are points
where the stretching is at maximum.

\subsection{Growth rate}

\begin{figure}[htb]
\centering
\makebox[8cm]{ \epsfxsize=7.0cm \epsfysize=6.0cm
\epsfbox{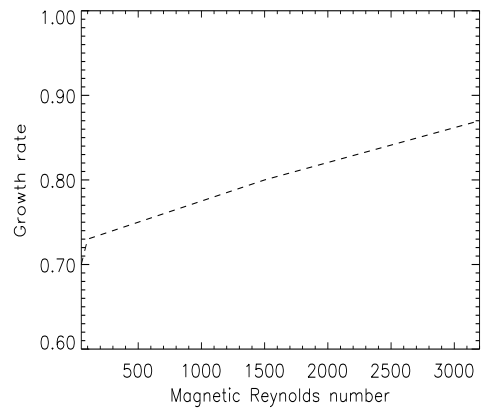}} \caption[]{\small Growth
rate of the magnetic energy versus the magnetic Reynolds number
for the $5:2:2$ flow.} \label{fig9}
\end{figure}

Numerical experiments were performed with ${\rm Re}_{\rm m}$
up to 3200 and numerical resolution up to $240^{3}$ to test
whether the $5:2:2$ case has vigorous dynamo properties in the
kinematic regime. The initial seed field was a random perturbation
with an amplitude of $10^{-5}$.  It was found that the magnetic
field is amplified exponentially and the growth rate increases
with ${\rm Re}_{\rm m}$ as illustrated in Fig.\
(\ref{fig9}).

\begin{figure}[htb]
\centering
\makebox[8cm]{ \epsfxsize=7.0cm \epsfysize=6.0cm
\epsfbox{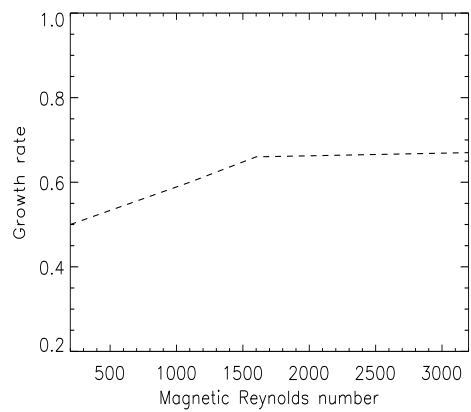}} \caption[]{\small Growth
rate of the magnetic energy versus the magnetic Reynolds number
for the $5:2:1$ flow.} \label{fig10}
\end{figure}

The $5:2:1$ case was also examined at high magnetic Reynolds
number. The highest value of ${\rm Re}_{\rm m}$ considered here is
${\rm Re}_{\rm m}=3200$. The growth rate seems to saturate at a
value of about 0.67 as illustrated in Fig.\ (\ref{fig10}).

\subsection{Flux ribbons}

As mentioned above, the $5:2:2$ flow has four points with minimum
velocity that coincide with local maxima of the function in Eq.\
(\ref{convergence.eq}).

\begin{figure}[htb]
\centering
\makebox[8cm]{ \epsfxsize=8.0cm \epsfysize=8.0cm
\epsfbox{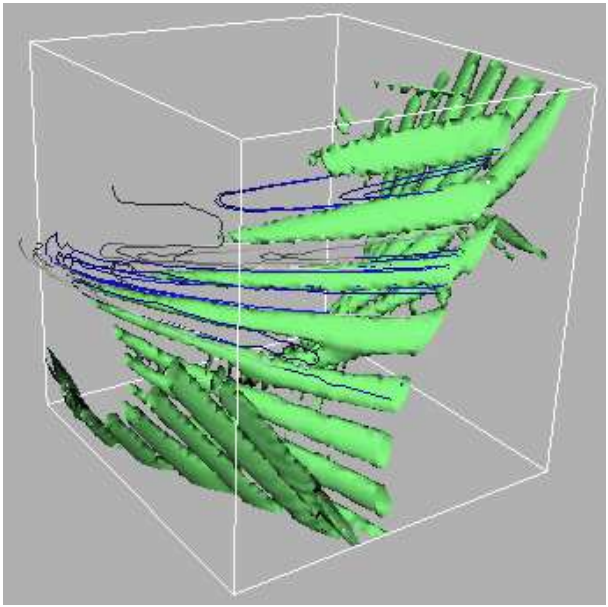} } \caption[]{\small The
magnetic field takes the form of flux ribbons when $A:B:C=5:2:2$.
The reconnection of field lines is shown when ribbons with
opposite polarity field are found close together.}
\label{fig11}
\end{figure}

When the induction equation is evolved from a weak random field,
flux ribbons arise close to the minimum velocity points. The
thickness of the ribbons decreases as ${\rm Re}_{\rm m}$
increases, presumably scaling as ${\rm Re}_{\rm m}^{-1/2}$,
similarly to the flux cigars for the standard ABC flow.  The sign
(polarity) of their magnetic field changes from ribbon to ribbon.
Thus, field lines with opposite polarity may reconnect when two
ribbons are close together. The magnetic flux is released through
the reconnected field lines (Fig.\ \ref{fig11}). However,
the magnetic ribbons are not confined at certain points but are
advected by the flow, and field lines with the same polarity could
also be found to approach each other.

\subsection{The amplification mechanism}

The process responsible for driving the amplification mechanism of
the mode with the largest growth is the constructive folding of
the ribbons.

Two of the most important properties of the $5:2:2$ flow is the
exponential stretching of the field lines and the way in
which the magnetic field is advected and the field lines are
packed.  Neighboring regions with magnetic field of opposite
polarity are destroyed whereas regions with like polarity flux are
conserved. Therefore the rate of field generation depends on the
global folding of the field lines.

\begin{figure}[htb]
\centering
\makebox[8cm]{ \epsfxsize=8.0cm \epsfysize=8.0cm
\epsfbox{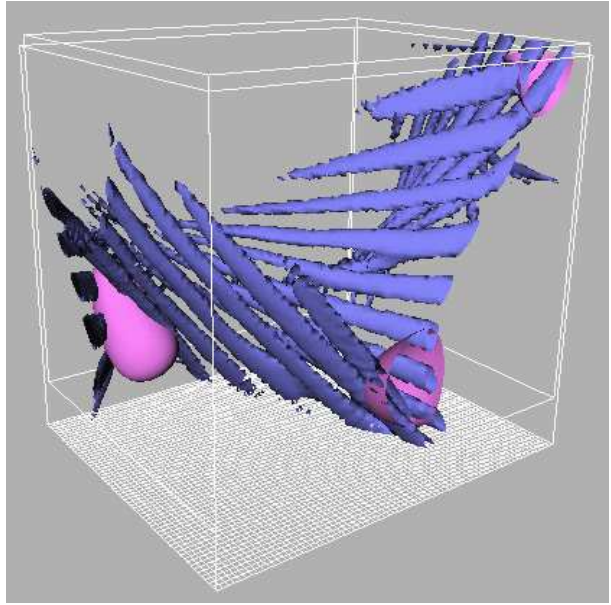} } \caption[]{\small A
snapshot showing the merge of the ribbons (dark) with like
polarity in the neighborhood of the minimum velocity points
(light).} \label{fig12}
\end{figure}

The flow folds the field so as to bring together flux in the same
direction. Thus, stretched flux ribbons with the same polarity
come into alignment and merge. The constructive folding of the
sheets occurs in the vicinity of the minimum velocity points
(Fig.\ \ref{fig12}). Due to the stretching ability of the flow,
which is maximum close to these points, the field lines are
stretched in these regions.

The flow advects the field so as to bring together stretched field
lines with the same polarity. They come into alignment and are
folded constructively. These field lines are continuously
stretched and also advected by the flow, and move to the vicinity
of another neighboring point with minimum velocity, where they
fold with new stretched field lines of the same polarity. The
folding is again constructive.

Due to the stretching and constructive reinforcement of the
magnetic field discussed above, the magnetic energy increases each
time the process is repeated.

\section{Conclusions}
\label{conclusions.sec}

The amplification of the dynamo generated field is
associated with a combination of stretching, twisting, folding and
reconnection of magnetic field lines. The importance of the weak
part of the field in the dynamo process is emphasized when the
wavenumber of the flow increases. Symmetry breaking allows
the transportation of stretched weak magnetic
field from cell to cell, where it gets packed against the local
flux structures (``cigars"). Most of the space in the bulk part of
the flow has passive advection, which does not change at increasing
magnetic Reynolds number. The evidence indicates that the invariant
properties of the flows (i.e.\ stretching ability) do not
change in the limit of arbitrarily small magnetic diffusivity.

The study of kinematic dynamo action in ABC flows is, from a
physical point of view, simple enough since it does not contain
the complexity of non-linear back-reaction of the magnetic field.
A fully non-linear study of the
dynamo problem connects the initial phase of the growth of a weak
magnetization state with the saturation of the dynamo generated
field.

We have demonstrated that the fast dynamo action reflects
the same mechanisms at work whatever the values of the parameters
(such as $\Rm$ and $k$) are. The main mechanism is the stretching of
the weak magnetic field. This is a first step towards the
understanding of the processes at work in a turbulent environment
and during the saturation state of a dynamo where the flow has
still a good grip on the weakest part of the magnetic field.

It is of course also of great interest to perform dynamo simulations
in more realistic (e.g. spherical) geometries, and to include more physical
processes such as differential rotation and convection. The
advantage of the simulations presented in the current paper is that the
individual mechanisms are easily identified and can studied in
great detail.
Thus, kinematic dynamo experiments such as these can provide very
useful clues for the general theory of astrophysical dynamos.

\begin{acknowledgements}
VA thanks the EU-TMR for support through a Marie Curie Fellowship.
SBFD was supported through an EU-TMR grant to the European Solar
Magnetometry Network. The work of {\AA}N was supported in part by
the Danish National Research Foundation, through its establishment
of the Theoretical Astrophysics Center. Computing time at the
UNI$\bullet$C computing center was provided by the Danish Natural
Science Research Council.
\end{acknowledgements}


\begin{thebibliography}{}

\bibitem[1983]{Arnold+ea83}
Arnold, V. and Korkina, E. 1983, Vest.\ Mosk.\ Un.\ Ta.\ Ser.\ 1,
Matem.\ Mekh., 3, 43

\bibitem[1996]{Cattaneo+ea96}
Cattaneo, F., Hughes, D.W., and Kim, E. 1996, Phys.\ Rev.\
Letters, 76, 2057

\bibitem[1995]{Childress+Gilbert1995}
Childress, S. and Gilbert, A.D.  1995, Stretch, Twist, Fold:The
Fast Dynamo (Springer-Verlag,Berlin,1995), 52

\bibitem[1986]{Dombre+ea86}
Dombre, T., Frisch, U., Green, J., Henon, M., Mehr, A., and
Soward, A.M 1986, JFM, 167, 353

\bibitem[2000]{Dorch2000}
Dorch, S.B.F. 2000, Physica Scripta, 61, 717

\bibitem[1993]{Galanti+ea93}
Galanti, B., Pouquet, A., and Sulem, P.L. 1993, in Theory of Solar
and Planetary Dynamos, Cambridge University Press, 99

\bibitem[1984]{Galloway+ea84}
Galloway, D. and Frisch, U. 1984, Geophy.\ \& Astroph.\ Fluid
Dyn., 29, 13

\bibitem[1984]{Galloway+ea86}
Galloway, D. and Frisch, U. 1986, Geophy.\ \& Astroph.\ Fluid
Dyn., 36, 53

\bibitem[1997]{Galsgaard+Nordlund1997}
Galsgaard, K. and Nordlund, {\AA}. 1997, Journ. Geophys. Res.,
102, 219

\bibitem[1979]{Hyman1979}
Hyman, J.M. 1979, Adv.\ Comp.\ Meth.\ for PDEs---III, Ed.\ R.\
Vichnevetsky, R.S. Stepleman, Publ.\ IMACS, 313

\bibitem[1993]{Lau+Finn93}
Lau, Y.-T. and Finn, J.M. 1993, Phys.Fluids B, 5, 365

\bibitem[1992]{Nordlund+ea92}
Nordlund, A., Brandenburg, A., Jennings, R.L., Rieutord, M.,
Roukolainen, J., Stein, R.F., and Tuominen, I. 1992, ApJ, 392, 647

\bibitem[1994]{Nordlund+ea94}
Nordlund,A.,Galsgaard,K., Stein,R.F.1994, in R.J.Rutten,C.J.
Schrijver(eds.) "Solar Surface Magnetic Fields", Vol.433,NATO ASI
Series

\bibitem[1972]{Vainshtein+Zeldovich1972}
Vainshtein, S.I. and Zeldovich, Ya.B. 1972, Usp.\ Fiz.\ Nauk, 106,
431

\bibitem[1993]{Zheligovsky+Pouquet1993}
Zheligovsky, O. and Pouquet, A. 1993, in Theory of Solar and
Planetary Dynamos, Cambridge University Press, 347

\end{thebibliography}
\end{document}